\documentclass[twocolumn,english,prl,amssymb,tightenlines,showpacs]{revtex4-1}
\usepackage[T1]{fontenc}
\usepackage[active]{srcltx}
\usepackage{amsmath}
\usepackage{amssymb}
\usepackage{graphicx}

\makeatletter
 
 \@ifundefined{textcolor}{}
 {%
   \definecolor{BLACK}{gray}{0}
   \definecolor{WHITE}{gray}{1}
   \definecolor{RED}{rgb}{1,0,0}
   \definecolor{GREEN}{rgb}{0,1,0}
   \definecolor{BLUE}{rgb}{0,0,1}
   \definecolor{CYAN}{cmyk}{1,0,0,0}
   \definecolor{MAGENTA}{cmyk}{0,1,0,0}
   \definecolor{YELLOW}{cmyk}{0,0,1,0}
 }

\makeatother

\usepackage{babel}
\begin{document}

\title{Direct observation of the degree of quantum correlations using photon-number
resolving detectors}

\author{L. Dovrat}

\affiliation{Racah Institute of Physics, the Hebrew University of Jerusalem, Jerusalem
91904, Israel}

\author{M. Bakstein}

\affiliation{Racah Institute of Physics, the Hebrew University of Jerusalem, Jerusalem
91904, Israel}

\author{D. Istrati}

\affiliation{Racah Institute of Physics, the Hebrew University of Jerusalem, Jerusalem
91904, Israel}

\author{E. Megidish}

\affiliation{Racah Institute of Physics, the Hebrew University of Jerusalem, Jerusalem
91904, Israel}

\author{A. Halevy}

\affiliation{Racah Institute of Physics, the Hebrew University of Jerusalem, Jerusalem
91904, Israel}

\author{H. S. Eisenberg}

\affiliation{Racah Institute of Physics, the Hebrew University of Jerusalem, Jerusalem
91904, Israel}

\begin{abstract}
Optical parametric down-conversion is a common source for the generation
of non-classical correlated photonic states. Using a parametric down-conversion
source and photon-number resolving detectors, we measure the two-mode
photon-number distribution of up to 10 photons. By changing the heralded
collection efficiency, we control the level of correlations between
the two modes. Clear evidence for photon-number correlations are presented
despite detector imperfections such as low detection efficiency and
other distorting effects. Two criteria, derived directly from the
raw data, are shown to be good measures for the degree of correlation.
Additionally, using a fitting technique, we find a connection between
the measured photon-number distribution and the degree of correlation
of the reconstructed original two-mode state. These observations are
only possible as a result of the detection of high photon number events.
\end{abstract}
\maketitle

Non-classical states of light are an essential resource for novel
protocols in quantum information and quantum
metrology\,\cite{Tapster1991,Dowling2008}. The most common tool
for producing such states is the nonlinear process of optical
parametric down-conversion (PDC). In this process, a parent pump
photon is split in a nonlinear material into two daughter
down-converted photons, while conserving energy and momentum. As
the down-converted photons originate from a single quantum system,
they possess correlations in many degrees of freedom, such as
their polarization, frequency and momentum. Because for any photon
emitted into one optical mode there is a sister photon emitted
into the other optical mode, there are also photon-number
correlations between the two modes. These correlations have been
used to produce heralded Fock
states\,\cite{Hong1986,WaksNumberStates2006,Mosley2008} and
enhance the precision of optical
measurements\,\cite{Rarity1987,Tapster1991,Eisenberg2005,Dowling2008}.

The down-converted photons are distributed over a range of spatial
and spectral modes. However, most experiments require that the
photons occupy a single mode. A specific mode is then
post-selected by spatial and spectral filtering of the photons.
The collected modes must be carefully matched to obtain a
high-quality produced state\,\cite{Wasilewski_Banaszek2008}. The
collection of matching modes would result in a joint photon-number
distribution of a non-classical highly-correlated state. On the
other hand, collecting two unrelated modes would result in a
classical joint photon-number distribution which is a product of
the two individual states. Recent developments in photon-number
resolving detectors allow the direct measurement of the joint
photon-number distribution and photon-number correlations between
two down-converted modes. However, imperfections in the detection
process, such as collection losses and false detections, alter the
measured photon statistics and reduce their correlations.
Non-classical correlations of multimode distributions from PDC
have been demonstrated indirectly using reconstruction
techniques\,\cite{Waks2004,Avenhaus2008} and directly with a
system of relatively low loss\,\cite{Waks2006,Allevi2010}

In this Letter, we use a novel photon-number resolving detection
scheme in order to measure the joint photon-number distribution of
a collinear type-II PDC process. These distributions were measured
up to the 10 photon terms. The two down-converted photons, which
have orthogonal polarizations, are each collected from a single
spatial and spectral mode. The degree of correlation was
controlled by varying the amount of overlap between the two
collected modes. We directly observe the transition between a
separable product state and a highly correlated state, despite the
presence of low detection efficiency and other distorting effects.
We introduce measures for the degree of correlations between the
two modes and relate these measures to the degree of
non-classicality of the collected state.

The PDC states are generated by a type-II collinear
$\beta-\textrm{BaB\ensuremath{_{2}}O\ensuremath{_{4}}}$ nonlinear
crystal. The crystal is pumped by amplified and frequency doubled
Ti:Sapphire laser pulses at a repetition rate of 250\,kHz. The
orthogonally polarized down-converted photons at 780\,nm are split
using a polarizing beam-splitter and coupled into separate silicon
photomultiplier (SiPM) photon-number resolving
detectors\,\cite{Bondarenko1998} (\emph{Hamamatsu Photonics},
S10362-11-100U). Before coupling to the detectors, the
down-converted photons are spatially filtered using single-mode
fibers and spectrally filtered using 3\,nm bandpass filters, to
ensure the collection of a single spatio-temporal
mode\,\cite{Dovrat2012}. The degree of correlation $g$ of the
measured photon state is determined by the amount of spectral and
spatial overlap between the two collected modes. The amount of
overlap is tuned by translating the optical fibers in order to
collect different spatial modes and by tilting the bandpass
filter, in order to shift its spectral band. The amount of overlap
is evaluated using the heralded efficiency $\gamma$, defined as
the ratio between the coincidence and the single count rates in
the limit when the average number of photons approaches zero. The
parameter $\gamma$ is linearly proportional to the degree of
correlation between the modes $g$ and to the overall photon
detection efficiency $\eta$. $\gamma$ was evaluated using standard
photon-number non-discriminating detectors (\emph{Perkin Elmer}
SPCM-AQ4C).

\begin{figure}
\noindent \begin{centering}
\includegraphics[width=1\columnwidth]{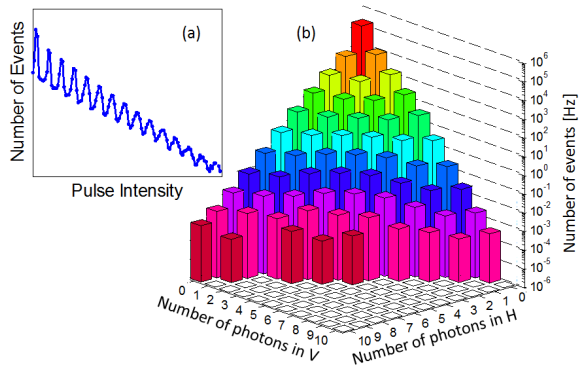}
\par\end{centering}

\caption{(color online) (a) Photon number resolution of a single
detector, demonstrated by the good peak separation of the pulse
intensity histogram. (b) The joint photon-number distribution of
two polarization modes (H and V) of a type-II collinear PDC
process for a maximally correlated state with
$\gamma=0.18$.\label{fig:18eff_bars}}
\end{figure}

The SiPMs are composed of a two-dimensional array of avalanche
photo diodes (APD) operating in Geiger mode. These detectors are
operable at room temperature and require a relatively low
operating voltage. They generate a current which is proportional
to the number of detected photons. The output signals are sampled
simultaneously within a 1\,ns sampling window, and analyzed in
real-time using programable electronics. A computer receives the
results and continuously displays the joint photon-number
distribution between the two polarization modes. The dark count
and afterpulsing rates\,\cite{Buzhan2006}, which generally limit
the number of resolvable photons, are minimized by synchronizing
the sampling time of the SiPM analog output with the arrival time
of the photons. Furthermore, the detectors are moderately cooled
using a thermoelectric cooler to $\sim-10^{\circ}$C. The bias
voltage is adjusted accordingly, so that the detection efficiency
is not affected. As a result, we obtain a good photon-number
resolution with an error of less than $1\%$. Figure
\ref{fig:18eff_bars}(a) shows an example of the histogram of the
output intensities, in which up to 14 photons can be resolved.

A measurement of the joint photon-number distribution between the
horizontal and vertical polarization modes with maximal
spatio-temporal overlap is shown in Fig.~\ref{fig:18eff_bars}(b).
We measured the full joint probability matrix up to 10-photon
terms. Ideally, the photon-number distribution of the correlated
state would be composed of only diagonal elements which correspond
to the same number of photons in both modes. However, the
distribution exhibits a large number of non-zero probability
values for events which contain different photon numbers. Similar
probability values were measured for events which involve the same
number of photons, indicated by the different color groups in
Fig.~\ref{fig:18eff_bars}(b). Note however that the probability of
events which contain no photons in either one of the modes are
slightly higher than the remaining probabilities within the same
group.

The distortion in the photon-number distribution is a result of
the imperfect detection process in SiPM
detectors\,\cite{Dovrat2012}. Not every photon impinging on the
detector creates a signal due to imperfect overall photon
detection efficiency. Additionally, false signals can be generated
by thermally-excited discharges (dark counts). Furthermore, when a
detection element is triggered, it might trigger additional
neighboring elements due to optical crosstalk\,\cite{Buzhan2006},
in which a spurious photon generated during a discharge in one APD
element propagates and is detected by another element. These
inherent effects alter the photon number distributions, reducing
the photon-number correlations even between highly correlated
modes.

\begin{figure}
\noindent \begin{centering}
\includegraphics[clip,width=1.0\columnwidth]{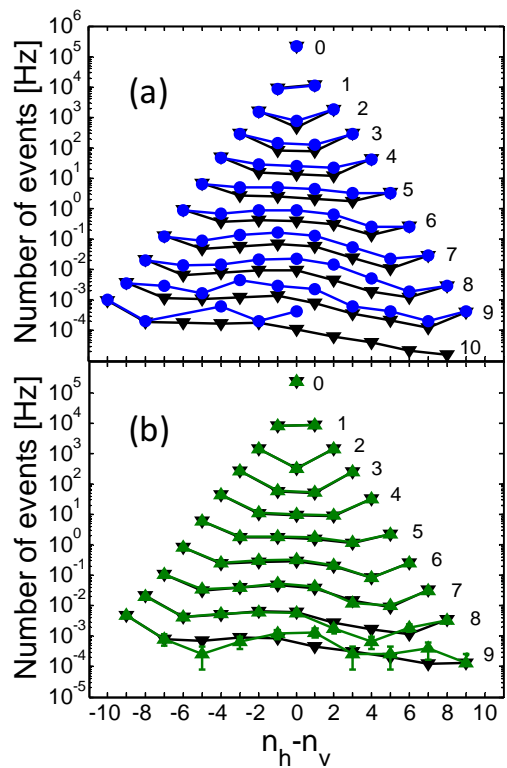}
\par\end{centering}

\caption{(color online) Joint distribution measurements between (a) highly
correlated modes $(\gamma=0.18$, blue circles) and (b) non-correlated
modes ($\gamma=0.06$, green up triangles). The corresponding products
of the two individual distributions are shown as black down triangles.
Poissonian errors are assumed and presented when larger than the symbol
size.\label{fig:diagonals_2D}}
\end{figure}

In order to study the photon-number correlations in our
measurements, we compare the measured correlated distribution with
its corresponding non-correlated product distribution, obtained by
multiplying the photon-number probabilities $P(n_{h})$ and
$P(n_{v})$ of the individual polarization modes. $n_{h}$ ($n_{v}$)
is the number of photons in the horizontal (vertical) polarization
mode. This result is shown in Fig.~\ref{fig:diagonals_2D}(a). The
distribution is displayed such that each curve connects events
which contain the same total number of photons $S=n_{h}+n_{v}$,
and the event counts are presented as a function of the photon
number difference $D=n_{h}-n_{v}$. This representation is similar
to that of the joint distribution in Fig.~\ref{fig:18eff_bars}(b),
if each color group were to be joined with a solid line. The
correlated distribution shows a clear distinction from its
corresponding product result. On the other hand, a similar
measurement between two relatively uncorrelated modes, results in
a distribution which is almost identical to its corresponding
product result (see Fig.~\ref{fig:diagonals_2D}(b)).

\begin{figure}
\noindent \begin{centering}
\includegraphics[width=1\columnwidth]{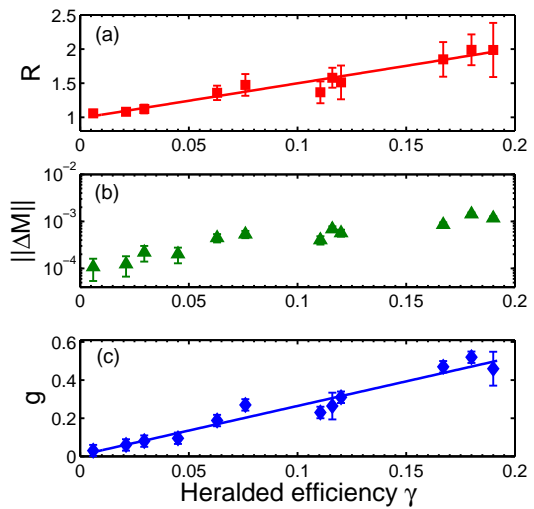}
\par\end{centering}

\caption{(color online) Quantitative measures for the degree of
correlation of the joint probability distributions. (a) The ratio
$R$ of the joint probabilities to their respective product values.
(b) The Euclidean distance $\left\Vert \Delta M\right\Vert $
between the measured distribution and its closest product state.
(c) The reconstructed degree of non-classicality
$g$.\label{fig:analysis}}
\end{figure}

We have recorded a series of joint probability distributions for
different values of the heralded efficiency $\gamma$. By
maintaining a fixed value for the overall detection efficiency,
the heralded efficiency can be used as a direct measure of the
mode overlap. In order to quantify the deviation of the joint
distribution from an uncorrelated product state, we define the
ratio between the joint photon-number probabilities and the
corresponding product of the probabilities of their individual
polarization modes
\begin{equation}
R(n_{h},n_{v})=P(n_{h},n_{v})/P(n_{h})\cdot P(n_{v}).\label{eq:R}
\end{equation}
As can be seen from Fig.~\ref{fig:diagonals_2D}(a), the values of
$R$ are fairly uniform for all probability values except for the
extremes, when there are zero photons in one of the modes. The
average values of $R$ obtained for all probabilities
$R(n_{h}\neq0,n_{v}\neq0)$ for the series of distributions are
presented in Fig.~\ref{fig:analysis}(a). These values approach
$R\approx1$ for the least correlated states and increase linearly
up to $R\approx2$ for the highly correlated state. Numerical
calculations of the dependence of $R$ on the degree of
non-classicality in the original state confirmed the linear
relation between the two parameters. The exact value of $R$
depends on the specific values of the average number of photons in
the original distribution, the detection probability, the dark
count rate and the crosstalk probability. Thus, even though it may
seem that the original correlations are completely washed out, the
ratio between the joint and product probabilities can be used as a
direct measure for the degree of correlation between the two
modes.

A mathematical tool which provides a more universal quantitative
measure for the similarity between a given two-mode distribution
and its closest product state is the singular value
decomposition~\cite{SVD_Article}. The decomposition of an $n\times
n$ probability matrix results in $n$ singular values $s_{i}$,
ordered as $s_{1}\geq s_{2}\geq\dots\geq s_{n}\geq0$ and
normalized to $\sum_{i}s_{i}^{2}=1$. The decomposition of a
separable product state results in a single non-zero value,
$s_{1}\neq0$, whereas a maximally correlated state results in $n$
values of $s_{i}=1/\sqrt{n}$. The normalized Euclidean distance
between a measured matrix $M$ and the closest product state
$M_{prd}$ can be expressed using the singular values of $M$ and is
given by $||\Delta M||\equiv\left\Vert M-M_{prd}\right\Vert
=\sqrt{s_{2}^{2}+s_{3}^{2}+\cdots+s_{n}^{2}}$.

The Euclidean distances $||\Delta M||$ for several heralded
efficiency values are shown in Fig.~\ref{fig:analysis}(b). The
errors in $||\Delta M||$ are estimated using a bootstrapping
procedure and assuming Poissonian noise. The distance exhibits a
clear increase by an order of magnitude as the heralded efficiency
is varied between its minimal and maximal values. Thus, the
Euclidean distance $||\Delta M||$ is a measure for the degree of
correlation of the two-mode state. It is applied directly to the
raw data, and can detect correlations despite large imperfections
in the detection apparatus.

In the experiment, the heralded efficiency $\gamma$ was first
optimized such that the largest amount of coincident events were
observed. This condition is fulfilled when the two collected
polarization modes maximally overlap, both spatially and
spectrally. Then, by misaligning one of the collected modes, the
overlap between the two modes was reduced, as well as the
photon-number correlations. Thus, the joint probability of the two
down-converted modes is a linear combination of the probability
$P_{pdc}$ of a correlated PDC distribution and that of an
uncorrelated product distribution $P_{prd}$
\begin{equation}
P(n_{h},n_{v},g)=g\cdot P_{pdc}(n_{h},n_{v})+(1-g)\cdot P_{prd}(n_{h},n_{v})\,,\label{eq:joint distribution}
\end{equation}
where the degree of correlation $g$ is the amount of overlap between
the two collected modes. For $g=1$, the distribution is that of a
collinear type-II PDC
\begin{equation}
P_{pdc}(n_{h},n_{v})=\begin{cases}
0 & n_{h}\neq n_{v}\\
\frac{1}{\left\langle n\right\rangle +1}\left(\frac{\left\langle n\right\rangle }{\left\langle n\right\rangle +1}\right)^{n} & n_{h}=n_{v}=n\,,
\end{cases}\label{eq:P_PDC}
\end{equation}
where the parameter $\left\langle n\right\rangle $ is the average
number of photons in each mode. For $g=0$, the distribution is that
of a product state of two thermal modes
\begin{equation}
P_{prd}(n_{h},n_{v})=\left(\frac{1}{\left\langle n\right\rangle +1}\right)^{2}\left(\frac{\left\langle n\right\rangle }{\left\langle n\right\rangle +1}\right)^{n_{h}+n_{v}}\,.\label{eq:P_Product}
\end{equation}

\begin{figure}
\noindent \begin{centering}
\includegraphics[width=1\columnwidth]{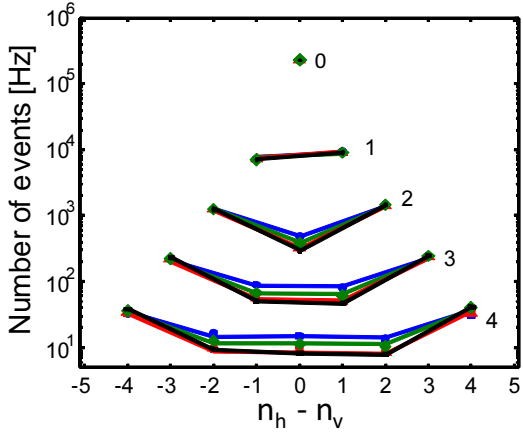}
\par\end{centering}

\caption{(color online) Measurements of distributions with fixed
product statistics and different heralded efficiencies of
$\gamma=0.17$ (clue circles), $\gamma=0.11$ (green diamonds),
$\gamma=0.02$ (red triangles) and the calculated product state
(black triangles). The solid lines are fits to Eq.~\ref{eq:joint
distribution} with the following parameters: average number of
photons $\left\langle n\right\rangle =4.1\pm0.1$, detection
efficiencies: $\eta_{A}=0.012\pm0.005$ and
$\eta_{B}=0.010\pm0.0005$, crosstalk probabilities:
$\epsilon_{A}=0.12\pm0.01$ and $\epsilon_{B}=0.11\pm0.01$, average
dark count rate: $\Delta_{A}=0.11\pm0.01$ and
$\Delta_{B}=0.14\pm0.01$.\label{fig:transition_fits}}
\end{figure}

By modelling the effects of loss, dark counts and crosstalk in the
SiPM detectors, we reconstructed the original two-mode
photon-number distribution using a fitting method, similar to that
presented in Ref.\,\cite{Dovrat2012}. The measured distribution
$P_{m}$ is related to the original distribution $P$ as
\begin{equation}
\bar{P}_{m}=\textrm{M}_{ct}\cdot\textrm{M}_{dk}\cdot\textrm{M}_{loss}\bar{P}\,,\label{eq:P_measured}
\end{equation}
where $\bar{P}_{m}$ and $\bar{P}$ are vector representations of
the measured and the original two-mode probability matrices,
respectively. The matrices $\text{M}_{loss}$, $\textrm{M}_{dk}$,
and $\textrm{M}_{ct}$ represent respectively the effects of loss,
dark counts and crosstalk according to the detector model of
Ref.\,\cite{Dovrat2012}. The fitting procedure is performed in two
stages. First, we perform a least squares fit to the computed
product state and obtain the average number of photons per mode
$\left\langle n\right\rangle $, the overall detection probability
$\eta$, the average number of dark counts, and the crosstalk
probability for each of the two modes. Then, the measured data is
fitted to Eq.~\ref{eq:P_measured}, with the degree of correlation
$g$ as the free parameter.

Some example results of the fitting process are shown in
Fig.~\ref{fig:transition_fits}. Three distributions taken for
different values of $\gamma$ are presented. The photon-number
probabilities in both polarization modes were kept constant for
all measurements, thus maintaining the same product state (black
lines). Transition from a highly correlated state to a product
state is observed as the value of the heralded efficiency is
decreased. The solid lines in Fig.~\ref{fig:transition_fits} are
fits to Eqs.~\ref{eq:joint distribution}. All fits result in
similar values for the dark counts, detection efficiency,
crosstalk probability, and $\left\langle n\right\rangle $ up to
the experimental error. The fits clearly differ in their values
for the degree of correlation $g$.

\begin{figure}
\begin{centering}
\includegraphics[width=1\columnwidth]{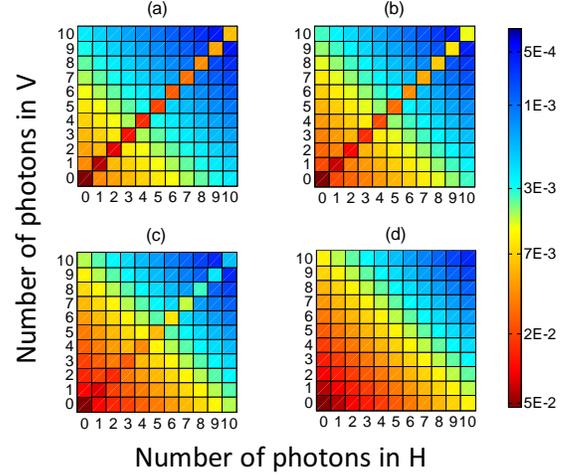}
\par\end{centering}

\caption{(color online) The reconstructed photon-number
probability distributions for the data presented in
Fig.~\ref{fig:transition_fits}. (a) $\gamma=0.17$, $g=0.47$ (b)
$\gamma=0.11$, $g=0.23$ (c) $\gamma=0.02$, $g=0.06$ and (d) the
product state.\label{fig:Reconstruction}}
\end{figure}

Figure~\ref{fig:Reconstruction} shows the reconstructed
distributions from the data of Fig.~\ref{fig:transition_fits}. For
the highest correlated state we observe strong photon-number
correlations, which gradually disappear as the value of the
heralded efficiency $\gamma$ is decreased. Even for the lowest
measured value of $\gamma$, photon-number correlations are still
evident.

The degree of correlation $g$ for additional values of $\gamma$
are shown in Fig.~\ref{fig:analysis}(c). As expected, the degree
of correlation $g$ depends linearly on the heralded efficiency.
The parameter $g$ is also the degree of non-classicality of the
state. Applying the non-classicality criterion for the photon
statistics of two-mode radiation of Lee\,\cite{Lee1990_2modes} on
Eq.~\ref{eq:joint distribution} shows that a necessary (but not
sufficient) condition for non-classicality is that $g>0$. We have
tested all of the reconstructed distributions against Lee's
criterion and found that all states with $g>0$ satisfy it.

In conclusion, we have measured the two-mode photon-number
distribution of a collinear type-II PDC process for different
degrees of correlation. Clear evidence for photon-number
correlations are presented despite the low detection efficiency,
the dark counts and the optical crosstalk effects of SiPM
number-resolving detectors, which highly distort the
number-correlations. These observations are only possible as a
result of the detection of high photon number events. Both the
ratio between the measured and the product matrix probabilities,
and the singular value decomposition of the measured probability
matrices, are shown to be good measures for the degree of
correlation. These two criteria are derived directly from the raw
data. Additionally, using a fitting technique, we found a
connection between the measured photon-number statistics and the
degree of correlation of the reconstructed original two-mode
state.
\begin{acknowledgments}
The authors thank O. Gat for fruitful discussions.
\end{acknowledgments}

\bibliographystyle{apsrev4-1}

\end{document}